\documentstyle[twocolumn,aps,psfig]{revtex}
\begin{document}
\draft
 
\twocolumn[\hsize\textwidth\columnwidth\hsize\csname @twocolumnfalse\endcsname
\title{
First-principles calculation of the thermal properties of silver
}
\author{Jianjun Xie,$^{1}$ Stefano de Gironcoli,$^{2}$
Stefano Baroni,$^{2,3}$ and Matthias Scheffler$^{1}$}
\address{
$^{1}$Fritz-Haber-Institut der Max-Planck-Gesellschaft,\\
Faradayweg 4-6, D-14195 Berlin-Dahlem, Germany }
\address{
$^{2}$SISSA -- Scuola Internazionale Superiore di Studi Avanzati and
\\ INFM -- Istituto Nazionale per la Fisica della Materia, \\ via Beirut
2--4, I-34014 Trieste, Italy
}
\address{$^3$CECAM -- Centre Europ\'een de Calcul Atomique et
Mol\'eculaire \\ ENS, Aile LR5, 6, All\'ee d'Italie, 69007 Lyon, France}
\maketitle

\begin{abstract} 
{\small
The thermal properties of silver are calculated
within the quasi-harmonic approximation, by using phonon dispersions
from density-functional perturbation theory, and the pseudopotential
plane-wave method. The resulting free energy provides predictions for
the temperature dependence of various quantities such as the
equilibrium lattice parameter, the bulk modulus, and the heat
capacity. Our results for the thermal properties are in good agreement
with available experimental data in a wide range of temperatures. As a
by-product, we calculate phonon frequency and Gr\"uneisen parameter
dispersion curves which are also in good agreement with
experiment.} 
\end{abstract}
]
\narrowtext
\section{Introduction}
\label{sec:level1}

The study of the temperature dependence of the properties of materials
requires a proper account of nuclear motions. Within the framework of
density functional theory (DFT) \cite{Review}, a major breakthrough in
this field has been opened by the introduction of {\it ab-initio}
molecular dynamics by Car and Parrinello \cite{Car85}. Far from the
melting point, however, a more conventional (and yet largely
unexplored in practice) approach based on lattice dynamics proves to
be both more accurate and computationally efficient. In the {\it
harmonic approximation}, the crystal free energy is calculated by
adding a {\it static} contribution---which is accessible to standard
DFT calculations---to a {\it dynamical} contribution which is
approximated by the free energy of system of harmonic oscillators
corresponding to the crystal vibrational modes (phonons). The latter
is nowadays conveniently calculated by using density-functional
perturbation theory (DFPT)~\cite{Baroni87,Gonze90}. In the {\it quasi-harmonic}
approximation \cite{Wette69,Maradu,Bruesch}, some anharmonic effects can
be accounted for by allowing phonon frequencies to depend on crystal
volume. Among other advantages, the (quasi-) harmonic approximation
allows an explicit account of quantum effects on nuclear motion, which
can be important below the Debye temperature. Furthermore, analysis of
the normal vibration modes and of their individual contribution to the
free energy can explicitly reveal the mechanism driving the thermal
expansion, phase transitions, and the crystal stability. The main
concern about the lattice dynamics method is the range of validity of
the quasi-harmonic approximation. Calculations based on various
semi-empirical models~\cite{Foiles89,Althoff93,Xu91,Barrera97} as well
as on first-principles
methods~\cite{Gonze90,Biernacki89,Pavone,Pavone98} demonstrate that the
quasi-harmonic approximation provides a reasonable description of the
dynamic properties of many bulk materials below the melting
point. Very recently, first-principles calculations on the thermal
expansion of some simple $s$-$p$ metals indicate that the treatment of
anharmonic effects at the quasi-harmonic level provides a remarkable
good description of the structural and elastic properties of these
materials up to their melting points~\cite{Quong97}.

In this paper we apply the quasi-harmonic approximation to the study
of the thermal properties of the $4d$ noble metal Ag, such as 
thermal expansion, heat capacity, and temperature
dependence of the bulk modulus. To this end, we first calculate the
phonon dispersion curves as functions of volume, by using
DFPT \cite{Baroni87}. Our results demonstrate that all these
quantities can be accurately predicted from the present parameter-free
method in a wide range of temperatures.

The paper is organized as follows. In Sec. II, we briefly outline our
computational framework, as well as some definitions concerning the
physical quantities we have investigated. The results of our
calculations are then presented and discussed in Sec. III. Finally, in
Sec. IV we give our conclusions.
 
\section{Theory}
\label{sec:level2}
\subsection{Equation of state and thermal expansion}

For a given temperature, $T$, and volume, $V$, the equilibrium state
of an extended system such as a crystal is determined by the condition
that the Helmoltz free energy, 
\begin{equation} F(V,T)=U-TS,
\label{eq:free-energy} \end{equation}
is at a minimum with respect to
variations of all possible {\it internal degrees of freedom}, such as
e.g. atomic arrangements and electronic states. Here $U$ and $S$
indicate the internal energy and entropy, respectively. The {\it
equation of state} of the system is obtained from
Eq.(~\ref{eq:free-energy}) by equating the pressure to minus the volume
derivative of the free energy: 
\begin{equation} p= -\left (
\frac{\partial F}{\partial V}\right)_{T}. \label{eq:eq-state}
\end{equation} 
In the quasi-harmonic approximation, $F$ is given by
\begin{eqnarray}
 F(V,T) &=& E(V)+F_{\rm vib}(\omega,T) \nonumber\\ &\equiv&
E(V)+k_{\rm B}T\sum_{{\bf q}}\sum_{j} {\rm ln} \left \{ 2\sinh \left
(\frac {\hbar \omega_{j}({\bf q})} {2k_{\rm B}T}\right)\right\},
\label{eq:quasi-harm} 
\end{eqnarray} 
where $E$ is the {\it static}
contribution to the internal energy---which is easily accessible to
standard DFT calculations, $F_{vib}$ represents the vibrational
contribution to the free energy, and $\omega_{j}({\bf q})$ is the
frequency of the $j$-th phonon mode at wave vector {\bf q} in the
Brillouin zone (BZ). Anharmonicity is explicitly, though
approximately, accounted for by allowing $E(V)$ to deviate from a
quadratic behavior and by letting the phonon frequencies depend on
volume. Since the temperatures considered here are well below the
electronic energy scale, the contribution of the electronic
excitations to the thermal expansion is negligible and is not included
in the present work. The {\it equation of state} (\ref{eq:quasi-harm})
can now be written in the form
\begin{eqnarray} p(V,T) &= &
-\frac{\partial E}{\partial V}- \frac {\partial F_{vib}} {\partial V}
\nonumber \\ &=& - \frac{ \partial E} {\partial V} + \frac{1}{V}
\sum_{{\bf q}}\sum_{j} \gamma_{j}({\bf q}){\cal E}(\omega_{j}({\bf
q})), \label{eq:eq-state-II} 
\end{eqnarray} 
where $\gamma_{j} ({\bf q})$ is the Gr\"uneisen parameter corresponding
to the $({\bf q},j)$ phonon mode, defined as: 
\begin{equation} \gamma_{j}({\bf
q})=-\frac{\partial \omega_{j}({\bf q})}{\partial V}\frac{V}
{\omega_{j}({\bf q})}, \label{eq:gruneisen} 
\end{equation} and ${\cal E}(\omega_{j}({\bf q}))$ is the mean 
vibrational energy of the $({\bf q},j)$ phonon given by 
 \begin{equation} 
 {\cal E} (\omega_{j}({\bf q}))
 =\hbar \omega_{j} ({\bf q}) \left [ \frac{1}{2} + \frac{1}{{\rm exp}
 (\hbar \omega_{j} ({\bf q}) / k_{B}T) - 1 }
 \right ].  \label{eq:ph-energy} 
 \end{equation}

The thermal expansion is obtained directly from the {\it equation of state}
(\ref{eq:eq-state-II}) and the volume thermal expansion coefficient 
is defined as
\begin{equation}
\alpha_{V}=\frac{1}{V}\left(\frac{\partial V}{\partial T}\right)_{p} \quad .
\end{equation}
The temperature dependence of the bulk modulus is obtained from 
\begin{eqnarray} B(T) &= &V\left( \frac{
\partial^{2}F} {\partial V^{2}}\right)_{T} \nonumber \\
&=&V\frac{\partial^{2} E}{\partial V^{2}}+V\left(\frac{\partial^{2}
F_{vib}(\omega,T)}{\partial V^{2}}\right)_{T}. \label{eq:bulk-modulus}
\end{eqnarray}

Due to anharmonicity, the heat capacity at constant pressure, $C_{p}$,
is different from the heat capacity at constant volume, $C_{V}$. The
former, which is what experiments determine directly, is proportional
to $T$ at high temperature, while the latter goes to a constant which
is given by the classical equipartition law: $C_V \approx 3Nk_{\rm
B}$, where $N$ is the number of atoms in the system. The relation
between $C_{p}$ and $C_{V}$ is \cite{Kittel} 
\begin{equation}
C_{p}-C_{V}=\alpha_{V}^2 (T)BVT, \label{eq:heat-cap} 
\end{equation}
and $C_{V}$ is given by
\begin{equation}
C_{V}=k_{B}\sum_{\bf q}\sum_{j}\left( \frac{\hbar\omega_{j}({\bf q})}
{2k_{B}T}\right)^{2}\frac{1}{\sinh^{2}(\hbar\omega_{j}({\bf q})/2k_{B}T)}
\end{equation}

\subsection{Computational details} The static total energy, $E(V)$,
and phonon frequencies, $\omega_j({\bf q})$, are calculated by using
DFT and DFPT respectively, within the local density
approximation \cite{Ceperley}. We use separable norm-conserving
pseudopotentials \cite{Kleinman,Gonze} together with a plane-wave basis
set up to a kinetic-energy cutoff of 55 Ry. Sums over occupied
electronic states are performed by the Gaussian-smearing special-point
technique \cite{gaussian,Monkhorst}, using 60 {\bf k} points in the irreducible
wedge of BZ. Phonon frequencies are calculated on a (444) regular mesh
and Fourier-interpolated in-between. This Fourier interpolation
amounts to including real-space inter-atomic force constants up to the
nineth shell of neighbors.

\section{Results} 
\begin{figure}
\centerline{\hbox{
\psfig{figure=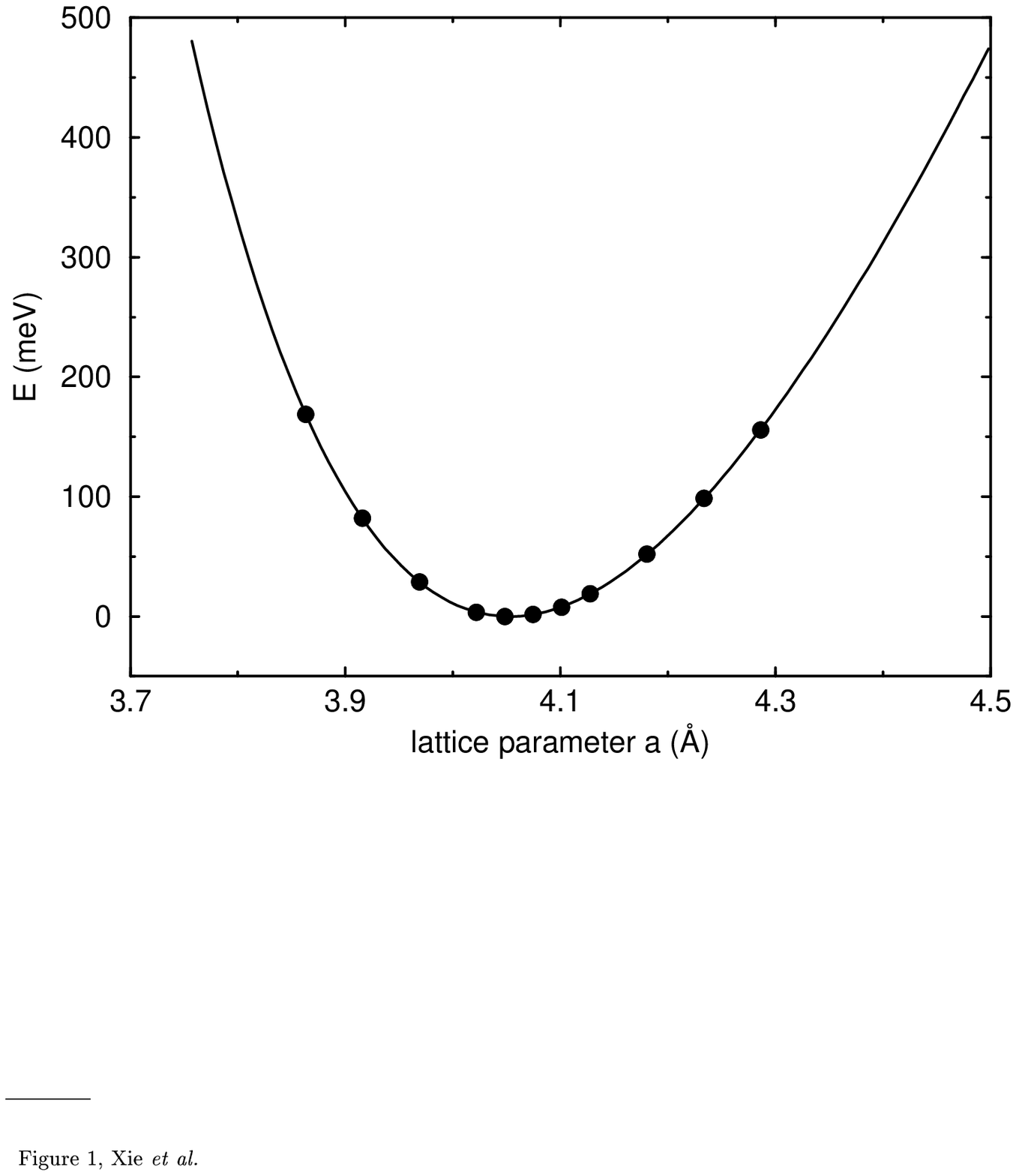,height=10cm}}}
\caption{Static total energy per atom as a function of lattice
parameter $a$} \label{fig:static-energy} \end{figure}

Figure \ref{fig:static-energy} shows the static
total energy per atom, $E(V)$, as a function of the lattice constant
$a=(4V)^{1\over 3}$ (Ag is a face-centered cubic metal). Our data are
fitted to a Murnaghan's equation of state \cite{Murna44}. The resulting
lattice constant, $a_{0}=4.05$~\AA, bulk modulus, $B=1.28$~Mbar,
and pressure derivative of the bulk modulus, ${\partial B/
\partial p} = 5.66$, agree well with previous theoretical
calculations \cite{Nara}. For comparison, the room-temperature
experimental data are: $a_0 = 4.08$~\AA~\cite{LANDOLT} and
$B = 1.01$~Mbar~\cite{Kittel}.

\begin{figure} 
\centerline{\hbox{
\psfig{figure=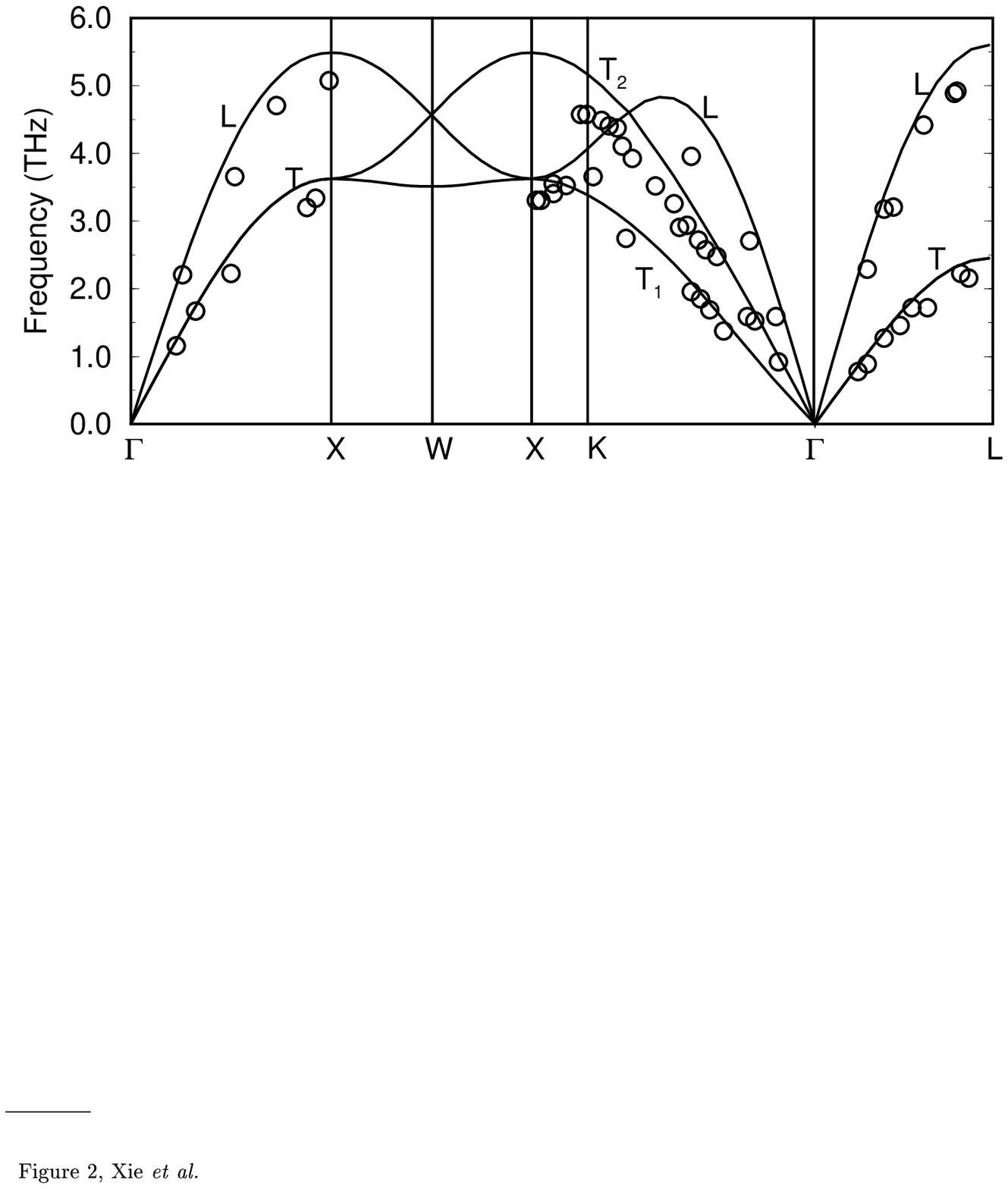,height=10cm}}}
\caption{Calculated phonon dispersion curves at the
lattice parameter corresponding to static equilibrium. Experimental 
neutron-scattering data~{\protect \onlinecite{Drexel}} are denoted by circles. 
T and L represent transverse modes and longitudinal modes respectively.}
\label{fig:phonons_eq}
\end{figure}

\begin{figure} 
\centerline{\hbox{
\psfig{figure=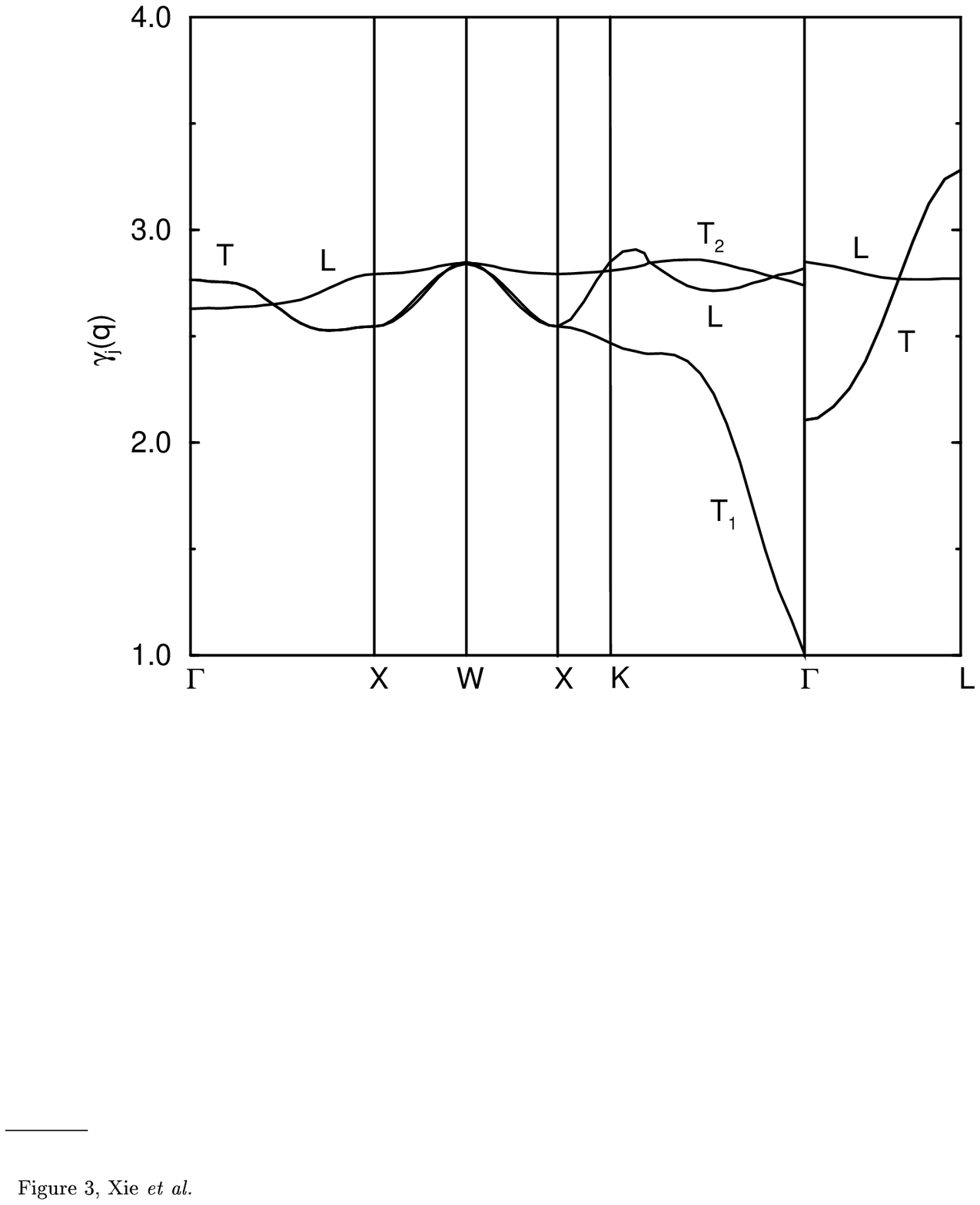,height=10cm}}}
\caption{Calculated dispersion curves of the mode
Gr\"{u}eisnen parameter $\gamma_{j}({\bf q})$ of silver along some
symmetry lines in BZ. T and L denotes the transverse modes and
longitudinal modes respectively.} \label{fig:gruneisen} \end{figure}

According to Eq. (\ref{eq:eq-state-II}), in order to obtain the
equation of state one must first calculate the phonon band structure
as a function of volume. In Figure \ref{fig:phonons_eq} we display 
the phonon dispersion curves as calculated along several symmetry
directions at the minimum of the static energy.
Experimental data at room temperature \onlinecite{Drexel} are reported 
for reference. The effect of temperature on the phonon dispersions will 
be discussed later in this paper. Figure~\ref{fig:gruneisen} shows the 
calculated dispersion curves of the mode
Gr\"{u}neisen parameters of Ag, as defined by
Eq.~(\ref{eq:gruneisen}), along the same symmetry directions. The
dispersions are discontinuous at the BZ center as a consequence of the
anisotropy and polarization dependence of the sound velocities. The
Gr\"{u}neisen parameters of silver are positive throughout the BZ for
all branches, thus implying that there is no anomalous negative
thermal expansion at low temperature, as in Si \cite{Ibach}. The
averaged Gr\"{u}neisen parameter of silver is 2.6, in agreement with
the experiment value of 2.5~\cite{Gschneider}.

\begin{figure} 
\centerline{\hbox{
\psfig{figure=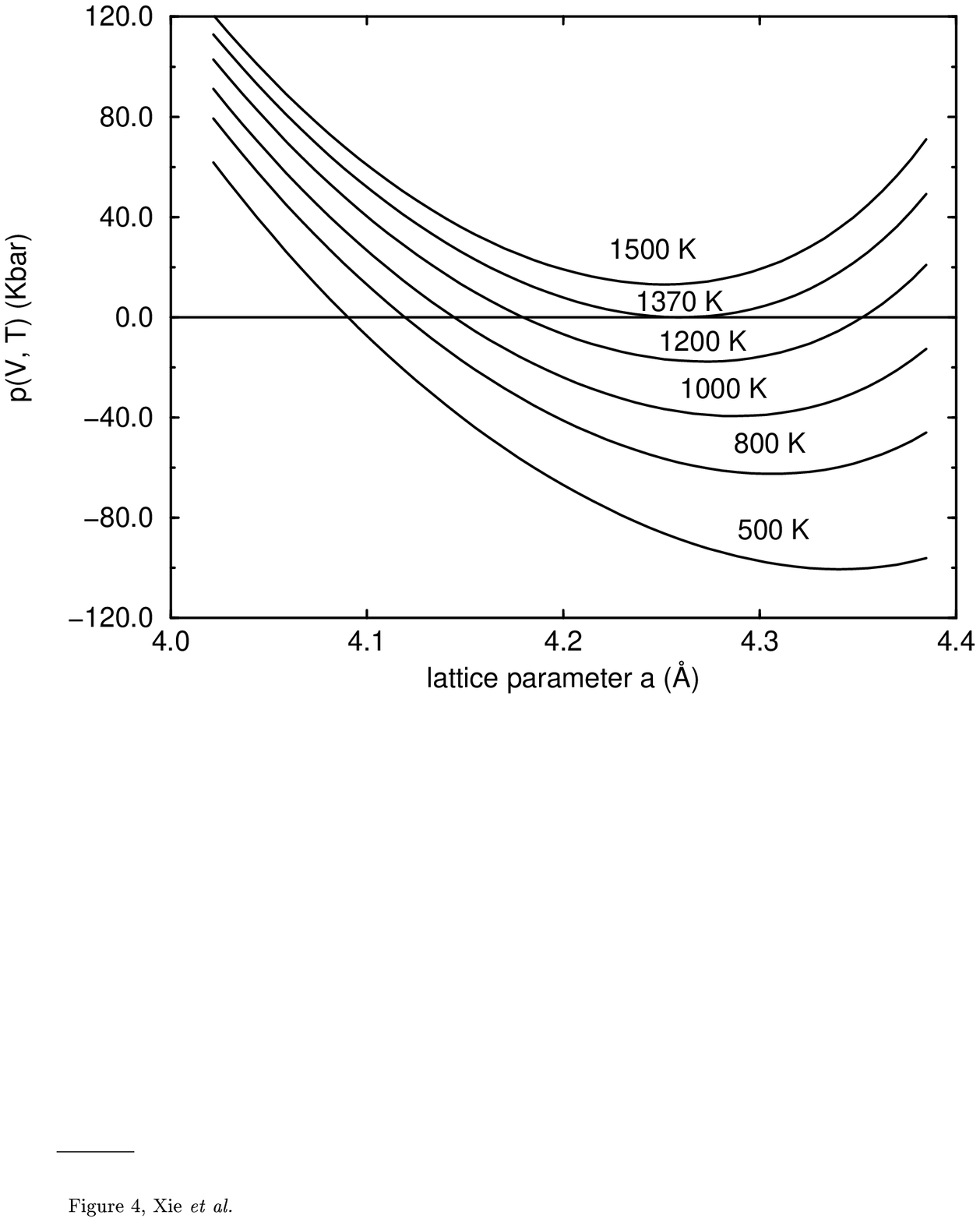,height=10cm}}}
\caption{Applied pressure $P(V,T)$ for silver as a
function of lattice parameter $a$ (here $V=a^{3}/4$) for different
temperatures.} \label{fig:pressure} \end{figure}

With the static total energy and the Gr\"{u}neisen parameters in our
hands, we can set up the {\it equation of state} via
Eq.~(\ref{eq:quasi-harm}). Figure \ref{fig:pressure} shows the
pressure $p(V,T)$ as a function of the lattice parameter $a$ for
several temperatures. It can be seen that for a given pressure, the
lattice constant increases with the temperature. At room temperature
and zero external pressure, the calculated lattice constant is
$a_{0}=4.07$ \AA, which is closer to the room temperature experimental
value of 4.08 \AA \cite{LANDOLT} than the result of 4.05 \AA\ derived
from the static total energy. As the temperature is increased, a
critical temperature $T_{\rm m}$ exists above which $p(V,T)$ no longer
intersects the $p=0$ line and, within the quasiharmonic approximation,
the crystal becomes mechanicaly unstable due to the vanishing of the 
isothermal bulk modulus $-V(\partial p/\partial V)_{T}$ at $T_{\rm m}$. 
In Silver this lattice instability occurs at 1370~K, not far from the
experimental melting temperature (1234 K) \cite{Kittel}.

\begin{figure} 
\centerline{\hbox{
\psfig{figure=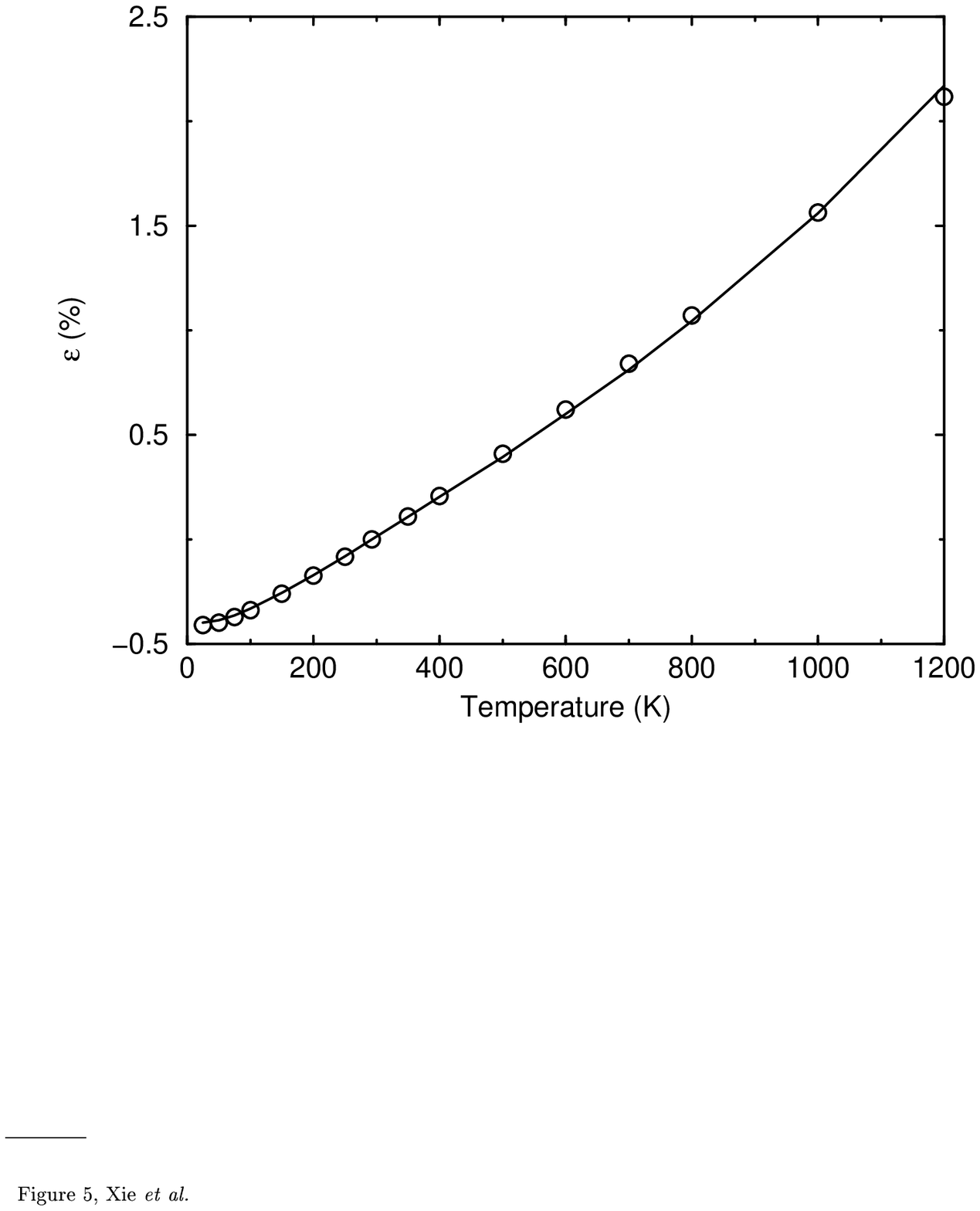,height=10cm}}}
\caption{Temperature dependence of the linear thermal
expansion for Ag. Solid curve is the calculated result and the circles
represent experimental data from Ref.~{\protect \onlinecite{Americ82}}.}
\label{fig:therm-expansion} \end{figure}

The thermal expansion can be derived directly from the {\it equation
of state}. We have calculated the linear thermal expansion which is
defined \cite{Americ82} as \begin{equation} \varepsilon = \frac{\Delta
a_{0}}{a_{0}}=\frac{a_{0}(T)-a_{0}(T=293)}{a_{0}(T=293)}.
\label{eq:therm-exp} \end{equation} The results are shown in Figure
\ref{fig:therm-expansion}. The agreement between the theoretical and
experimental results is very good. Although the calculated equilibrium
lattice constant is slightly different from experimental measurement,
the temperature dependence of the relative volume changes are
described accurately by the present method. The good agreement between
theory and experiment holds not only at low temperature, but also near
the melting point. 

\begin{figure} 
\centerline{\hbox{
\psfig{figure=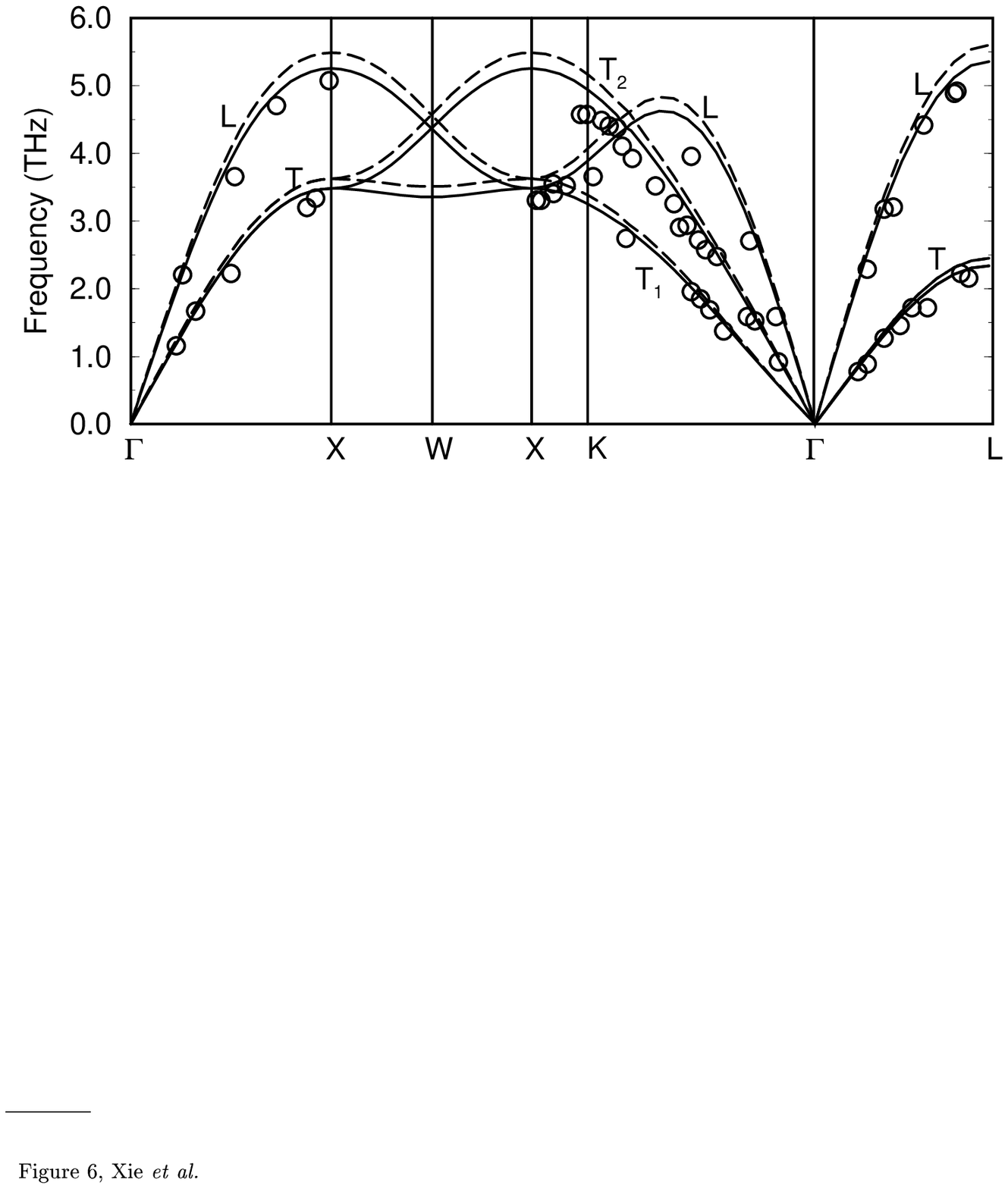,height=10cm}}}
\caption{Solid lines: calculated phonon dispersion curves of 
silver at $T= 293$ {\bf K}. Dashed lines: calculated phonon dispersions at the
lattice parameter corresponding to static equilibrium. Experimental 
neutron-scattering data~{\protect \onlinecite{Drexel}} are denoted by circles. 
T and L represent transverse modes and longitudinal modes respectively.}
\label{fig:phonons_roomtemperature}
\end{figure}

Given the volume as a function of temperature, the
temperature dependence of phonon frequencies can be approximately
estimated through their volume dependence: ${\bigl ( \omega(T) -
\omega(T=293) \bigr ) \over \omega(T=293) } \approx 3\gamma \varepsilon$,
where $\gamma$ is the Gr\"uneisen parameter (Eq.(~\ref{eq:gruneisen}))
and $\varepsilon$ is the linear thermal expansion,
(Eq.(~\ref{eq:therm-exp})). In the present case, given the fact that
$\varepsilon(T=0)\approx -5\times 10^{-3}$, and that Gr\"uneisen parameters
are tipically $\gamma \approx 2 \sim 3$, phonon frequencies at room 
temperature are lower than their $T=0$ value by $3 \sim 5 \%$.
Figure \ref{fig:phonons_roomtemperature} shows the theoretical phonon 
dispersions at room temperature (full lines) together with the dispersions,
which are already shown in Figure \ref{fig:phonons_eq} and 
calculated at the static 
equilibrium (dashed lines in Figure \ref{fig:phonons_roomtemperature}). 
The comparison with experimental data at room temperature (open circles) 
\cite{Drexel} is clearly improved and now rather satisfactory.

\begin{figure} 
\centerline{\hbox{
\psfig{figure=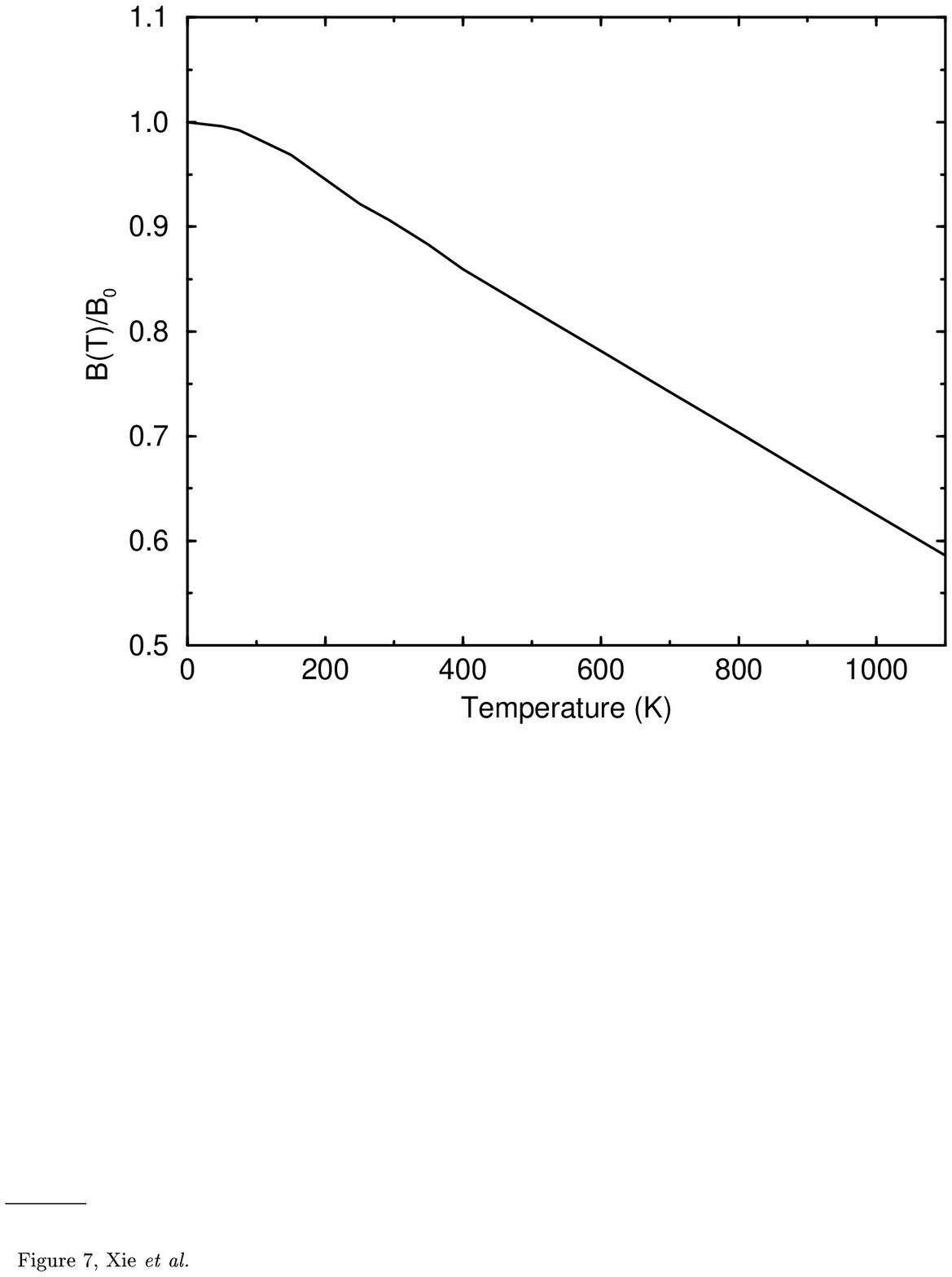,height=10cm}}}
\caption{Calculated temperature dependence of the ratio
of bulk modulus $B(T)$ to $B_{0}$ (obtained from static total energy)
for silver.} \label{fig:bulk} \end{figure}

Figure \ref{fig:bulk} shows the temperature dependence of the bulk
modulus as calculated from Eq.(~\ref{eq:bulk-modulus}). $B_{0}$ denotes
the bulk modulus obtained from the static total energy by neglecting
the lattice vibrations. At room
temperature, we obtain a bulk modulus $B = 1.16 $ Mbar. Comparing
this value with the result derived from static LDA calculations
($B_{0}=1.28 $ Mbar), one sees that the agreement with the
experimental result ($B=1.01 $ Mbar) \cite{Kittel} is significantly
improved.

\begin{figure} 
\centerline{\hbox{
\psfig{figure=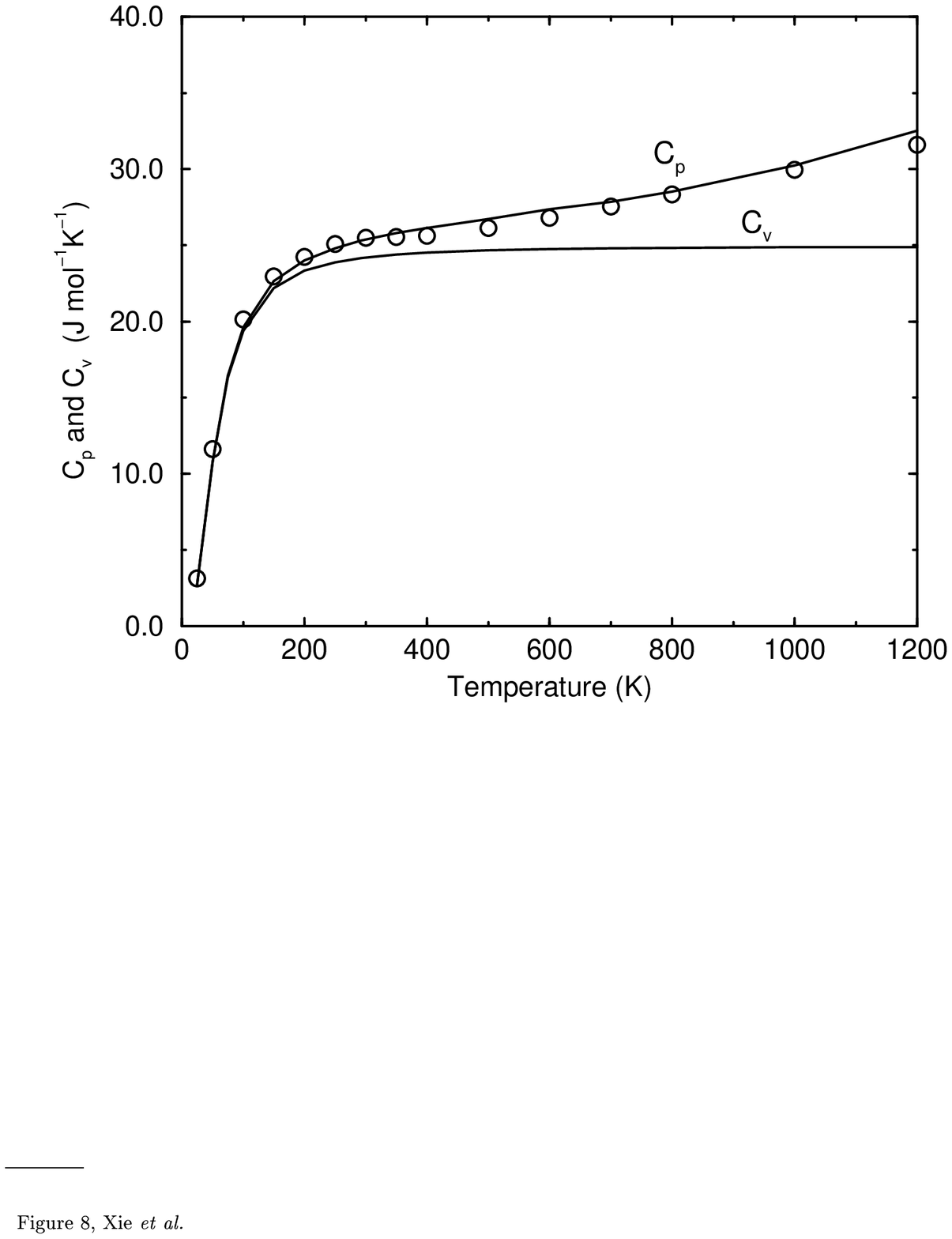,height=10cm}}}
\caption{ Calculated temperature dependence of heat
capacity of Ag at constant pressure ($ C_{p}$ ) and at constant volume
($C_{V}$). The experimental data for $C_{p}$~{\protect
\onlinecite{Americ63}} are denoted by circles.} \label{fig:heat-capacity}
\end{figure}

The calculated heat capacity $C_{p}$ and $C_{V}$ are shown in Figure
\ref{fig:heat-capacity}. It can be seen that below the Debye
temperature ($\theta_{D}=215 $ K for silver \cite{Gschneider}) the
difference between $C_{p}$ and $C_{V}$ is very small, while at high
temperature, the heat capacity at constant volume $C_{V}$ approaches
to the classical value 2.49 Jmol$^{-1}{\rm K}^{-1}$, while the heat
capacity at the constant pressure increases monotonously with the
temperature. The available experimental data \cite{Americ63} for
$C_{p}$ are shown as circles. The agreement between theory and
experiment is remarkable in a wide range of temperatures also in this case.

\section{Conclusions} In the present paper, we have calculated the
thermal properties of silver, such as thermal expansion coefficient,
Gr\"{u}neisen parameters, bulk modulus, and heat capacity, using the
the quasi-harmonic approximation within density-functional theory. The
equilibrium lattice constant is obtained from the {\it equation of
state} constructed by the free energy. The volume dependence of
frequencies is calculated from the density-functional perturbation
theory. The obtained results for the investigated thermodynamic
quantities are in good agreement with the available experiment
measurements. The calculation suggests that the anharmonic properties
of silver can be accurately calculated from this first-principles
approach in a wide range of temperature. The application of the
present method to the study of other properties of materials such as
high pressure effect, crystal stability and phase transitions, is 
straightforward.

\acknowledgments

One of the authors (J.J. Xie) would like to acknowledge the financial
support from Alexander von Humboldt foundation in Germany.
Two of us (SB and SdG) have done this work in part within the 
{\sl Iniziativa Trasversale Calcolo Parallelo} of INFM.

\newpage

\end{document}